\documentstyle[prl,aps]{revtex}

\begin{document}

\newcommand{\rr}{{\bf r}}
\newcommand{\la}{\langle}
\newcommand{\ra}{\rangle}
\newcommand{\Om}{\Omega}
\newcommand{\be}{\begin{equation}}
\newcommand{\ee}{\end{equation}}
\newcommand{\ba}{\begin{eqnarray}}
\newcommand{\ea}{\end{eqnarray}}
\newcommand{\half}{\frac{1}{2}}
\newcommand{\quarter}{\frac{1}{4}}
\newcommand{\eighth}{\frac{1}{8}}
\newcommand{\R}{{\bf R}}
\newcommand{\Q}{{\bf Q}}
\newcommand{\x}{{\bf x}}
\newcommand{\qr}{{\Q \cdot \R}}
\newcommand{\xiR}{\xi(\R)}
\newcommand{\etaR}{\eta(\R)}
\newcommand{\DxiR}{\Delta\xiR}
\newcommand{\DZR}{\Delta Z(\R)}
\newcommand{\p}{{\rm Pr}}

\title{Inversion of Randomly Corrugated Surfaces Structure from
Atom Scattering Data}

\author{Daniel A. Lidar}
\address{
Chemistry Department\\
University of California\\
Berkeley, CA 94720\\
USA  
}

\maketitle

\begin{abstract}
The Sudden Approximation is
applied to invert structural data on randomly corrugated surfaces from
inert atom scattering intensities. Several expressions relating
experimental observables to surface statistical features are
derived. The results suggest that atom 
(and in particular He) scattering can be used profitably to study
hitherto unexplored forms of complex surface disorder.
\end{abstract}

\section{Introduction}

Structurally disordered surfaces have been a subject of great interest
for some time now. Of special interest are epitaxially grown films,
liquid surfaces, and amorphous surfaces.  In epitaxial growth for
example, metal or semiconductor atoms are adsorbed on a surface under
thermal conditions, to form two- and three-dimensional structures on
top of it. The physical and chemical properties are determined by the
final form of these structures. These may be of dramatic importance,
e.g, in the production of electronic devices. One of the most exciting
aspects of epitaxial growth kinetics, is that it {\em {prepares
disordered structures}} in the intermediate stages. The disorder
manifests itself in the formation of various types of clusters or
diffusion limited aggregates on top of the surface. These structures
may be monolayers (usually at high temperatures, when the diffusivity
is large, or at coverages significantly below a monolayer), in which
case the disorder is two-dimensional, or they may be composed of
several layers, giving rise to disorder in three
dimensions. Epitaxially grown structures of this type offer an
exceptional opportunity for both experimental and theoretical study of
disorder. No satisfactory and comprehensive theory of the epitaxial
growth process is as of yet available, much due to the absence of
reliable interaction potentials for the system. The situation with
respect to liquid and amorphous surfaces is similar: very little is
known at this point about their structure. Progress at this stage thus
hinges critically on data available from experiments. An important
experimental technique is thermal atom scattering, and in particular
He scattering
\cite{Engel:82,Benny:review2,me:CAMP}.
The main advantage offered by He scattering is complete {\em surface}
sensitivity, as He does not penetrate into the bulk, unlike other
scattering techniques such as neutron or X-ray scattering, or
low energy electron diffraction (LEED). Another important advantage is
that He scattering is highly 
non-intrusive, due to the inertness and low mass of the He atoms. The
latter also means that He scattering is really a {\em diffraction}
experiment at the typical meV energy scale at which most experiments
are performed, with sensitivity to {\em atomic-scale} features.  The
interpretation of He scattering experiments is, however, rather
involved due to the complicated interaction between the He beam and
the surface. As in all other scattering problems, this interpretation
issue is in fact one of {\em inversion} of the He/surface
potential. The inversion problem, 
however, is intrinsically ill-posed, since one can only measure {\em
intensities}, 
not phases. There is a certain redundancy in the intensity information
which can be exploited to obtain {\em relative} phase shifts, but
never the {\em absolute} phases. Thus the inversion can never be fully
performed, although
partial information can be obtained, or useful approximations can be
made concerning the shape of the potential, which yield an
analytically closed-form solution
\cite{Benny:inversion1,Rabitz:inversion1,me:optical}.

In this work I will
concentrate on the problem of trying to relate surface {\em structural
features} to the scattering intensities, rather than the potential in
general. This involves a non-trivial step of connecting the potential
with such features. One possibility is the suggestion by Norskov and
coworkers \cite{Norskov} on the basis of
effective-medium theory, in which the leading repulsive term in the He
scattering potential from any electronic system is taken to be
proportional to the local unperturbed electron density $n_0({\bf r})$
of the host at the He position ${\bf r}$:

\begin{equation}
V({\bf r}) = \alpha n_0({\bf r}) .
\label{eq:Norskov}
\end{equation}

\noindent In this way the potential is rather simply related to
structural features such as the local electronic corrugation. However,
while useful for describing the interaction close to the surface,
where repulsion dominates, this formula is inapplicable at large
distances where the He/surface interaction is dominated by the
long-range attractive forces. In addition, it still requires knowledge of the
local electron density, often a highly nontrivial task, especially in
presence of defects or impurities. Thus in practice one often resorts 
to the use of specific functional forms for the potential, such as the
Morse or Lennard-Jones potentials, and fits the parameters to the
experimental data. Unfortunately, the connection to surface
geometrical features is then less transparent.

So far almost all of the work done on inverting He/surface potentials
has been for He scattering 
from {\em ordered}, crystalline surfaces. Again, a full inversion is
impossible, but useful results have been obtained by assuming specific
forms for the interaction potential. These include semiconductor
surfaces: GaAs(110), Si, InSb(110)
\cite{Laughlin:82}, and some transition metal
surfaces (Ni, Cu, Ag, Au)
\cite{Garcia:83}. More recent
efforts have concentrated on potentials for scattering from various
adsorbed monolayers on metals, e.g., the c(2$\ast$2) phase of oxygen
on Ni(001) \cite{Godby:85}, hydrogen-plated Pt \cite{Schwartz:86}, and
Xe/Cu(110) \cite{Ramseyer:94}. We recently fitted cross-section data
to obtain a potential for a disordered Ag/Pt(111) system
\cite{me:Ag-systems}. A very detailed recent review of potentials of
physical {\em adsorption} is the work of Vidali {\it et al.}
\cite{Vidali:91}, which tabulates parameters of interest as deduced
from analysis of experimental data and calculations of over 250
gas-surface systems, including He.

Formal inversion methods have also occupied the attention of various
researchers. The first such work was presented by Gerber and Yinnon
\cite{Benny:inversion1}, who showed that atom-surface interaction
potentials can be recovered from the diffraction peak intensities
measured in beam scattering experiments by a direct, simple inversion
method, using the Sudden approximation (SA). The SA is a highly
successful and useful theoretical method in the He scattering field,
and has been reviewed by Gerber
\cite{Benny:review2}. The method of
Ref. \cite{Benny:inversion1} was applied to simulated Ne scattering
from W(110). This was followed by the first inversion of real
atom/surface scattering data: the He/MgO(100) system
\cite{Benny:inversion2}. Schlup \cite{Schlup:89} showed how the
surface profile function can be inverted in the Eikonal approximation,
and applied his method to simulated data. In Ref. \cite{me:optical} we
studied a related problem: the inversion of an ad-atom profile
function in the SA. Rabitz and coworkers developed a method based on
functional sensitivity analysis and Tikhonov regularization
\cite{Viswanathan:90}, and applied this to the inversion of the He
potential for scattering from a Xe monolayer on the (0001) face of
graphite \cite{Rabitz:inversion1,Rabitz:inversion2}, which was also
attempted earlier by employing close-coupling calculations
\cite{Hutson:inversion}. Finally, we recently inverted the structure
of a low temperature disordered overlayer of Ag on Pt(111)
\cite{me:SSL97}.

It should be noted that the situation with regards to potentials for
He scattering from disordered surfaces is very much inferior to the
case of ordered surfaces described above. Essentially no such reliable
potentials exist, and the subject is at this point at a most
preliminary stage. The common approach it to assume that the potential
can be represented as a surface term plus a sum of pairwise additive
terms representing the interaction of the He with each of the surface
adatoms or vacancies. In this paper I will take a different approach,
and will show how one can derive statistical information about
randomly corrugated surfaces from He scattering measurements. This
information comes in the form of correlation functions, not potential
parameters. The resulting expressions relate useful statistical
parameters characterizing the surface disorder to experimental
observables which are not hard to obtain in practice, such as
incidence energy dependence of the specular intensity. Unfortunately,
at the time of writing experiments are unavailable for comparison with
the results obtained here. The developments will therefore be
primarily methodological, in anticipation of experimental data. It is
hoped that the results obtained here will motivate He scattering
experiments on disordered solid and liquid surfaces. As demonstrated
by this work and others before it, He scattering can provide a wealth
of information on disordered surface structure and dynamics.

The structure of the paper is as follows. Sec. \ref{SA} provides a
brief introduction into the SA. Sec. \ref{info} is the heart of the
paper and derives the inversion expressions. Concluding remarks are
brought in Sec.\ref{conclusions}.

\section{Brief Review of the Sudden Approximation}
\label{SA}

Consider a He atom with mass $\mu$ incident upon a surface with
wavevector ${\bf k} \!=\! ({\bf K},k_z)$. $\hbar {\bf K} \!=\!
\hbar(k_x,k_y)$ and $\hbar {\bf K}'$ are respectively the intial and
final momentum components parallel to the surface, and $\hbar k_z$ is
the incident momentum normal to the surface. The position of the He
atom is ${\bf r} \!=\! ({\bf R},z)$, where ${\bf R} \!=\!  (x,y)$ is
the lateral position. The SA is valid when the collisional momentum
transfer ${\bf q} \!=\! {\bf K}'-{\bf K}$ in the direction parallel to
the surface is much smaller than the momentum transfer normal to the
surface: $2 k_z \gg |{\bf q}|$. This condition is satisfied at
relatively high incidence angle and energy $E \!=\! (\hbar {\bf
k})^2/(2m)$, and moderate surface corrugations. When it holds, one can
approximately consider the scattering along $z$ as occurring at fixed
${\bf R}$. Then if $\psi$ is the He wavefunction, it satisfies a
Schr\"{o}dinger equation where the dependence on ${\bf R}$ is {\em
adiabatic}:

\be
\left[ -{\hbar^2 \over 2\mu} {d^2 \over dz^2} + V_{\bf R}(z)
\right] \psi_{\bf R}(z) = \varepsilon \psi_{\bf R}(z) .
\label{eq:SE}
\ee

\noindent Here $V_{\bf R}(z)$ is the He-surface interaction potential
and no inelastic channels are included, so that the total energy
$\varepsilon$ is conserved . This means that each surface
point ${\bf R}$ gives rise to an elastic real phase shift $\eta({\bf
R})$, which can be evaluated in the WKB approximation from
Eq.(\ref{eq:SE}) as:

\ba
\eta({\bf R}) = \int_{\xi({\bf R})}^{\infty} dz\:
\left[\left( {k_z}^2 - {2m \over \hbar^2}V_{\bf R}(z) \right)^{1/2} -k_z
\right] - k_z\,\xi({\bf R}) ,
\label{eq:eta}
\ea

\noindent where $\xi({\bf R})$ is the classical turning point
pertaining to the integrand in Eq.(\ref{eq:eta}). The phase shift in
turn yields the S-matrix as: ${\cal S}({\bf R}) \!=\! \exp[2i \eta({\bf
R})]$. The ${\bf R}$ coordinate is conserved in this picture so the 
S-matrix is diagonal in the coordinate representation:

\[
\langle {\bf R}'| {\cal S} | {\bf R} \rangle = e^{2i \eta({\bf R})}
\delta({\bf R}'-{\bf R}) .
\]

\noindent Experimentally one measures probabilities $|{\cal
S}({\bf K}' \rightarrow {\bf K})|^2$ for ${\bf K}
\rightarrow {\bf K}'$ transitions. To obtain these $\langle
{\bf R} | {\bf K} \rangle \!=\! \exp(i {\bf K} \!\cdot\! {\bf
R})/\sqrt{A}$ (where $A$ is the area of the surface) can be used, to find:

\ba
\langle {\bf K}'| {\cal S} | {\bf K} \rangle =
\int d{\bf R}' \: d{\bf R}
\langle {\bf K}'|{\bf R}' \rangle
\langle {\bf R}'| {\cal S} | {\bf R} \rangle 
\langle {\bf R} |{\bf K}  \rangle =
{1 \over A} \int d{\bf R} e^{-i {\bf q} \cdot {\bf R}}
e^{2i \eta({\bf R})} .
\label{eq:Sud}
\ea

\noindent This is the well-known expression for the SA scattering
amplitude \cite{Benny:Sud1}. The SA has been tested
extensively by comparison to exact
coupled-channel calculations on Ne/W(110) and He/LiF(001)
\cite{Yinnon:78}, as well as by comparison to exact time-dependent
propagation methods in the case of scattering from defects
\cite{Benny:RBs}. It is particularly noteworthy that the SA yielding
an {\em inverted} potential of
remarkable accuracy \cite{Benny:inversion1} for
simulated Ne/W(110) data. The most important deficiency of the SA
(apart from being limited to high energies) is its inability to
describe double collision events. This is because, as mentioned above,
the ${\bf R}$ coordinate is conserved in the SA, i.e., each
trajectory takes place at constant ${\bf R}$. Clearly, no double collisions
can occur under such conditions. However, double collisions may take
place, e.g., when an incident atom is scattered off a defect onto the
surface, or in the opposite order. This issue was resolved recently
by combining the SA with the Born approximation \cite{me:SS98}.

\section{Inversion of Structure of Randomly Corrugated Surfaces from
H\lowercase{e} Scattering Intensities} 
\label{info}

\subsection{Atom Scattering from a Randomly Corrugated Square Well
Potential} 
\label{sudden}

Recall the role of $\xiR$ in the SA amplitude [Eq.(\ref{eq:Sud})]: it
is the position of the classical turning
points, which can alternatively be viewed as the surface corrugation
function. In the
case of a disordered surface $\xiR$ can be a random function of $\R$ and
in order to obtain observable quantities one must average over an
appropriate ensemble which characterizes the physical and statistical
properties of the surface of interest. Thus the scattering
probability, as a function of momentum transfer $\Q$, is given by

\begin{equation}
P(\Q) = \langle |S(\Q)|^2 \rangle ,
\label{eq:P}
\end{equation}

\noindent where $\langle \cdots \rangle$ indicates an average over the
ensemble of which $\xiR$ is a typical sample. In certain cases one
will be justified in assuming that translational invariance has been
established after averaging over the corrugation ensemble. For
example, this will generally be the case for liquid surfaces, for
solids when the structural disorder on the surface is due to radiation
damage, and epitaxially grown defects on a surface. This means that
$\xiR$ is a stationary stochastic process, and hence that $\langle
f[\xiR] \rangle$ is independent of $\R$, and that $\langle
g[\xiR,\xi(\R')] \rangle$ depends only on $\R\!-\!\R'$. The assumption
of stationarity will, however, not hold for the case in which the
structural disorder is caused by an incomplete adsorbed overlayer on a
periodic substrate, and I will present a different treatment for such
cases in a later section. Combining Eqs.(\ref{eq:Sud}),(\ref{eq:P}) one
obtains in the case of a translationally invariant potential

\begin{equation}
P(\Q) = {1 \over A} \int d\R \: e^{i \qr} \langle e^{2i
[\etaR-\eta(0)]} \rangle .
\label{eq:P-T.I.}
\end{equation}

\noindent From the above expression it can be seen that within the SA,
the angular scattering intensity is the Fourier transform of a
function of the random variable

\begin{equation}
E(\R) = 2[\eta(0)-\etaR] .
\label{eq:Eta}
\end{equation}

\noindent To proceed, one must first establish the connection between
the phase-shift and properties of the surface. Assume now that the
interaction of the He with the entire surface can be expressed in the
form of a square well potential of depth $\epsilon$. Square well
potentials have been rather successful in predicting properties of
interacting gases; the accuracy obtained in fitting second and third
virial coefficients with a square well potential compares to that of a
Lennard-Jones 6-12 potential \cite{Reichl}. They have also been
studied by others in the context of inversion problems
\cite{Chadan:89,Klibanov:92}, e.g., in neutron reflectometry from
magnetic films \cite{Haan:96}. In our case the potential 
assumes the form:

\[
{2m \over \hbar^2}V(\R ,z) = \left\{ \begin{array}{ll}
	\infty \:\:     & \mbox{: $z<\xiR$} \\
	-\epsilon \:\:  & \mbox{: $\xiR<z<\xiR+\DxiR=\zeta(\R)$} \\
	0 \:\:          & \mbox{: $z>\zeta(\R)$} \\
	\end{array} .
\right.
\]

\noindent Typically, $\xiR$ is a relatively strongly corrugated
function as it 
expresses the interaction of the He atom with the core electrons responsible
for the steep repulsive part of the potential. On the other hand,
$\zeta(\R)$ 
may be very smooth, reflecting the loss of detail at the long distances at
which the attractive part of the He-surface potential becomes
important. According to Eq.(\ref{eq:eta}) the phase shift is then given by

\begin{equation}
\eta(\R) =  k_z\,\DxiR \left( \left( 1+ \delta \right)^\half -1
\right) -k_z\,\xiR ,
\label{eq:eta-SW}
\end{equation}

\noindent where $\delta \!=\! {\epsilon/k_z^2}$ is a small parameter in the
high-energy SA. Eq.(\ref{eq:eta-SW}) is known as ``the Beeby effect''
\cite{Beeby}, i.e., in the presence of a well the wave 
number has to be replaced by an effective wave number which is due to
the acceleration of the particle by the attractive forces. In the case
of scattering by a pure hard wall [$\DxiR\!=\!0$], or a square well
with the same shape as the hard wall [$\DxiR\!=\!z_0={\rm constant}$],
the scattering intensity is

\begin{equation}
P(\Q) = {1 \over A} \int d\R \: e^{i \bf{Q \cdot R}} \langle e^{2i\,k_z
[\xi(0)-\xiR] } \rangle .
\label{eq:P-T.I.-hard}
\end{equation}

\noindent In this case the intensity is essentially the Fourier transform of
the {\em characteristic function} $\exp[i k_z \,Z(\R)]$ of the
relative surface corrugation:

\[
Z(\R) = 2[\xi(0)-\xiR] ,
\]

\noindent which makes its interpretation particularly clear. Knowledge
of the probability density $f_Z(z;\R)$ fully determines the
corrugation of the surface in question, in that $f_Z(z;\R) dz$ is the
probability of finding a corrugation of height between $z$ and
$z\!+\!dz$ at $\R$, with respect to that at the origin. It is a well
known fact in probability theory \cite{Mises} that for any (piecewise-)
continuous distribution function $f_X(x)$ a unique and explicit
inversion of $f_X(x)$ exists from the characteristic function. Hence
it is clear at the outset that He scattering can, at least in
principle, provide very useful information about the statistics of a
disordered system. In the following sections I will show how the
statistical information pertaining to the disordered surface can be
extracted from the He scattering intensity.

\subsection{Extraction of the Surface Corrugation Probability Density
from the 
Angular Scattering Intensity for a Hard Wall Potential}
\label{probability}

In this section I specialize to the hard wall potential. This model
has been in use for many years in He scattering theory
\cite{Flytzanis:74,Harris:83,Benny:Sud3} and is related to the Eikonal
approximation in optics \cite{Born}. In Ref. \cite{Harris:83} the
origin of the 
hard-wall was discussed and it was concluded that it is due to the the
local density of metal 
electron states in the selvedge. This leads to the
Esbjerg-Norskov theory, Eq.(\ref{eq:Norskov}). The great merits of the
hard wall approximation 
are that (1) it is analytically tractable and (2) it provides a direct
geometrical interpretation of the surface corrugation. However, it is
clearly oversimplified and leaves out many interesting features of the
He-surface interaction. More sophisticated approximation schemes have
therefore been introduced by various researchers. The {\it
distorted-wave Born approximation} was used early on
\cite{Weare:74} to treat {\em soft} potentials
\cite{Garcia:82} and low-energy scattering 
\cite{Armand:82}. These studies have shown significant deviations from
the predictions of the hard-wall model, e.g., for a corrugated Morse
potential for He scattering by a Cu(110) surface
\cite{Armand:82}. Nevertheless, the hard wall approximation has
physical merit in the He high-energy limit, which is assumed here in
connection with the SA.

An inverse Fourier
transform of the scattering intensity $P(\Q)$ [Eq.(\ref{eq:P-T.I.-hard})]
yields the characteristic function of the surface corrugation:

\begin{equation}
\langle e^{i k_z\,Z(\R)} \rangle = {A \over 2\pi} {\cal
F}^{-1}_{k_z}[P(\Q); \R] ,
\label{eq:characteristic}
\end{equation}

\noindent where a notation for the Fourier transform was introduced:

\begin{eqnarray*}
{\cal F}\left[f(x); \: y\right] = {1 \over \sqrt{2\pi}}
\int_{-\infty}^{\infty} dx\: f(x)\, e^{i \, x\, y} .
\end{eqnarray*}

\noindent In the case of normal incidence $P(\Q)$ is a symmetric
function, so 
that its Fourier transform is real. The probability density $f_Z(z;\R)$ can
now be found if one recalls that:

\begin{equation}
\langle e^{i k_z\,Z(\R)} \rangle = \int dz\: e^{i k_z\,z} f_Z(z;\R) .
\label{eq:characteristic1}
\end{equation}

\noindent Formally, therefore, $f_Z(z;\R)$ is fully determined by the
scattering intensity, through another Fourier transform:

\begin{equation}
f_Z(z;\R) = {A \over {(2\pi)^2}} \int dk_z\: e^{-i k_z\,z} {\cal
F}^{-1}_{k_z}[P(\Q); \R] .
\label{eq:f-inversion}
\end{equation}

\noindent The last equation shows that {\em He scattering may in
principle be 
used to fully invert the probability density of the corrugation
function of a 
randomly corrugated surface}. However, this requires dense sampling of the
scattering intensities over a broad range of incidence wave-numbers, a task
which may be difficult to accomplish in practice. Additional
difficulties may arise due to Gibbs phenomenon and noise. A more realistic
approach is 
to determine moments of the random variable describing the surface
corrugation, an idea which dates back to the Backus and Gilbert work
in geophysics \cite{Backus:67}, and has been pursued by others as well
\cite{Chadan:89,Louis:96}. To show how this may be accomplished here
one may expand the 
characteristic function in terms of its moments. We then obtain from
Eqs.(\ref{eq:characteristic}),(\ref{eq:characteristic1}):

\begin{equation}
\sum_{n=0}^{\infty} {(i k_z)^n \over n!} \langle Z(\R)^n \rangle = {A \over
2\pi} {\cal F}^{-1}_{k_z}[P(\Q); \R] \equiv g(k_z; \R) .
\label{eq:moments}
\end{equation}

\noindent The moments of $Z(\R)$ may now be found by differentiation of the
experimentally available function $g(k_z; \R)$:

\begin{equation}
\langle Z(\R)^n \rangle = {1 \over i^n}{{d^n g(0; \R)} \over {d
k_z^n}} .
\label{eq:moments-n}
\end{equation}

\noindent Thus in contrast to the rather involved task of extraction of the
full probability density via Eq.(\ref{eq:f-inversion}), requiring an
experiment to be performed over the entire energy range, extraction of the
moments merely involves an extrapolation of the data to the low energy limit
$k_z \!=\! 0$. Progressively higher moments, however, require higher
derivatives and hence an increasingly dense sampling in $k_z$ to
reduce noise 
due to finite differences. As a check, one may estimate the first moment
$\langle Z(\R) \rangle \!=\! 2\langle \xi(0)\!-\!\xiR \rangle$, which must
clearly vanish. In practice, perhaps the most interesting piece of
information 
is the second moment. Assuming that all higher moments vanish is
equivalent to 
the assumption that $Z(\R)$ is a Gaussian random variable. From
Eqs.(\ref{eq:moments}),(\ref{eq:moments-n}) one has:

\begin{equation}
\langle Z(\R)^2 \rangle = -{A \over 2\pi} {{d^2} \over {d
k_z^2}} \left[ {\cal F}^{-1}_{k_z}[P(\Q); \R] \right]|_{k_z=0} .
\label{eq:moments-2}
\end{equation}

\noindent One may also choose to focus on the corrugation function
$\xiR$ itself, instead of on the relative corrugation $Z(\R) \!=\!
2[\xi(0) \!-\!
\xiR]$. Without loss of generality one can take $\langle \xi(0)
\rangle \!=\! 
\langle \xiR \rangle \!=\!  0$. A cumulant expansion can then be used
to evaluate 
the averages in Eq.(\ref{eq:P-T.I.-hard}). Truncating this expansion
at the second cumulant, or equivalently, assuming that $\xiR$ is a
Gaussian random variable, one obtains:

\begin{equation}
P(\Q) = {1 \over A} \int d\R \: e^{i \bf{Q \cdot R}}
e^{(2\sigma\,k_z)^2 [C(\R)-1]} ,
\label{eq:P-cumulant}
\end{equation}

\noindent where the variance and correlation function are respectively

\begin{equation}
\sigma^2 = \langle \xiR^2 \rangle, \:\:\:\: C(\R)={1 \over \sigma^2} \langle
\xi(0) \xiR \rangle .
\label{eq:sigma-C}
\end{equation}

\noindent Here, by definition, $0\!\leq\! |C(\R)| \!\leq\! 1$ and $C(\R)
\rightarrow 0$ as $R \rightarrow \infty$. Fourier transforming the
scattering 
intensity now yields an expression which can be used to conveniently fit
$\sigma^2$ and $C(\R)$ as a function of $k_z$:

\begin{equation}
e^{(2\sigma\,k_z)^2 [C(\R)-1]} = {A \over 2\pi} {\cal
F}^{-1}_{k_z}[P(\Q); \R] .
\label{eq:P-cumulant1}
\end{equation}

\noindent A set of equations [notably
Eqs.(\ref{eq:f-inversion}),(\ref{eq:moments-2}),(\ref{eq:P-cumulant1})]
has thus been derived which
can be applied to extract useful statistical information on the surface
corrugation, within a hard wall model, from the angular intensity
distribution. In the next section I will consider what information can be
derived within the more realistic square well model.

\subsection{Correlation Functions in the Square Well Model}
\label{SW}

It is not possible to analytically solve for the probability density of the
effective surface corrugation or its moments in the square well model,
as was 
done in Sec.\ref{probability} in the hard wall case. It is, however,
possible 
to obtain some interesting information by performing a cumulant
expansion, as 
I now proceed to show. An inverse Fourier transform of the scattering
intensity in the general case of a translationally invariant potential
[Eq.(\ref{eq:P-T.I.})] yields [using Eq.(\ref{eq:Eta})]:

\[
\langle e^{i E(\R)} \rangle = {A \over 2\pi} {\cal F}^{-1}[P(\Q); \R] .
\]

\noindent Introducing the notation

\[
\DZR = 2[\DxiR - \Delta \xi(0)] .
\]

\noindent one obtains with the help of Eq.(\ref{eq:eta-SW}), after
some algebra:

\[
E(\R) \approx k_z \left[ Z(\R) -\half \delta \,\DZR \right] .
\]

\noindent This expression is exact to first order in the small parameter
$\delta$. Without loss of generality again set $\langle Z(\R) \rangle \!=\!
0$, and also $\langle \DZR \rangle \!=\! 0$, so that the first moment of
$E(\R)$ vanishes. For the second moment one obtains, again to first order in
$\delta$:

\[
\langle E(\R)^2 \rangle - \langle E(\R) \rangle^2 \approx {k_z}^2 \left[
\langle Z(\R)^2 \rangle - \delta \langle Z(\R) \,\DZR \rangle \right] .
\]

\noindent In analogy to Eq.(\ref{eq:sigma-C}) define a variance and
correlation function between the hard wall corrugation and the
deviation from 
it due to the attractive well:

\[
\sigma_1^2 = \langle \xiR\,\DxiR \rangle, \:\:\:\: C_1(\R)={1 \over
\sigma_1^2} 
\langle \xiR\,\Delta\xi(0) \rangle ,
\]

\noindent so that

\[
\langle Z(\R) \,\DZR \rangle = 8 \sigma_1^2 [1-C_1(\R)] .
\]

\noindent Collecting the results we obtain finally to second order in the
cumulant expansion:

\[
e^{(2\sigma\,k_z)^2 [C(\R)-1] - (2\sigma_1\,\epsilon)^2 [C_1(\R)-1]} = {A
\over 2\pi} {\cal F}^{-1}[P(\Q); \R] .
\]

\noindent This equation [compare to Eq.(\ref{eq:P-cumulant1})] may be
used to 
fit the experimental data to obtain the correlation functions,
variances, and 
the well depth of the attractive part of the potential. This is much
information to fit to a single experimental function. However, since the
contribution due to the attractive well is expected to be small, one may at
first neglect this contribution. This is tantamount to assuming a hard wall
interaction, by which one may obtain an estimate of the hard wall quantities
$\sigma$ and $C(\R)$. Having computed these, one may correct the fit by
inclusion of $\sigma_1$, $C_1(\R)$, and $\epsilon$, and proceed iteratively.
In the next section I will consider how statistical
information can be derived from the experimentally straightforward
measurement 
of the {\em specular} intensity.

\subsection{Extraction of the Correlation Length from Specular Peak
Measurements in the Hard Wall Model} 
\label{specular}

From an experimental point of view, there is a significant advantage to
working with specular intensities, as their measurement involves a
relatively 
minor effort compared to that required to obtain the full angular intensity
distribution. There is also a theoretical advantage, in that energy transfer
to the surface is significantly reduced in specular collisions, so that a
phononless treatment is more justified. In the hard wall approximation,
by addition and subtraction of the same term, Eq.(\ref{eq:P}) for the
angular 
distribution can be written as

\begin{eqnarray}
\lefteqn{
P(\Q) = \left| {1 \over A} \int d\R \: e^{i \qr} \langle e^{-2i k_z \xiR}
\rangle \right|^2 + } \nonumber \\
&&
{1 \over A^2} \int d\R \int d\R' \:  e^{i \Q \cdot (\R-\R')} \left[ \langle
e^{-2i k_z [\xiR-\xi(\R')]} \rangle - \langle e^{-2i k_z \xiR} \rangle
\langle 
e^{2i k_z \xiR} \rangle \right] . \nonumber
\end{eqnarray}

\noindent Assuming translational invariance again and specializing to the
specular direction, one obtains with a second order cumulant expansion (see
Ref. \cite{Benny:Sud3} for a more detailed derivation)

\begin{equation}
I \equiv P(0) = e^{-\beta E} \left[1+{1 \over A} \int d\R \:
\left(e^{\beta E 
C(\R)} -1 \right) \right] ,
\label{eq:specular}
\end{equation}

\noindent where for brevity:

\[
\beta = (2\sigma)^2, \:\:\:\:\: E=k_z^2 .
\]

\noindent (It was not sufficient to simply set $\Q=0$ in
Eq.(\ref{eq:P-cumulant}) since then the integral would diverge due to
the asymptotic properties of $C(\R)$.)

I now wish to evaluate the specular intensity $I$ for some specific
correlation functions. To be consistent with the truncation of the
cumulant expansion at second order, I will investigate the case of
short-range order, in which higher order moments of the corrugation
function do not play a significant role. One is thus led to consider
two types of short-range correlation functions. These are of course
just convenient models, assumed for lack of further knowledge of the
probability density distributions. In both cases I will
be concerned with evaluating the integral term in
Eq.(\ref{eq:specular}), denoted by:

\[
F(E) = {1 \over 2\pi} \int d\R \: \left(e^{\beta E C(\R)} -1 \right) .
\]

\subsubsection{Gaussian Correlation Function}
\label{Gaussian}

Let us assume a Gaussian form for the correlation function:

\[
C(R) = e^{-(R/l)^2} ,
\]

\noindent where $l$ is the correlation length. Then due to cylindrical
symmetry 

\[
F(E) = \int_0^L \left[ e^{\beta E \left(e^{-(R/l)^2}-1\right)} -1 \right] R
\,dR ,
\]

\noindent where $L$ is the linear extent of the surface. As such the
integral cannot be evaluated analytically, but its derivative with
respect to the energy can:

\begin{equation}
{{\partial F} \over {\partial E}} = {l^2 \over 2E} \left( e^{\beta E} -
e^{\beta E e^{-(L/l)^2}} \right) .
\label{eq:dF/dE-G}
\end{equation}

\noindent Since short-range order was assumed one may safely neglect
$\exp \left(-(L/l)^2 \right)$. Differentiating Eq.(\ref{eq:specular}) and
combining with Eq.(\ref{eq:dF/dE-G}) then yields: 

\[
{{\partial I} \over {\partial E}} + \beta I + {{\pi l^2} \over A}
{{e^{-\beta 
E} - 1} \over E} = 0 .
\]

\noindent This equation can be used conveniently for a best-fit of the
correlation length $l$ and the variance $\sigma^2 = \beta/4$, by using the
experimental {\em specular} data, as a function of incidence energy
$E=k_z^2$. 

\subsubsection{Exponential Correlation Function}
\label{exponential}

Next assume a longer range, exponential form for the correlation function:

\[
C(R) = e^{-(R/l)} ,
\]

\noindent where $l$ is a new correlation length. Then:

\[
F(E) = \int_0^L \left[ e^{\beta E \left(e^{-(R/l)}-1\right)} -1
\right] R\,dR .
\]

\noindent Again the integral cannot be evaluated analytically, but its
derivative can:

\begin{equation}
{{\partial F} \over {\partial E}} = {l^2 \over E} \left[ {\rm Ei}(\beta E) -
\log(\beta E) -\gamma \right] .
\label{eq:dF/dE-exp}
\end{equation}

\noindent [neglecting $\exp(-L/l)$]. Here ${\rm Ei}(x) \!=\! e^x \left(
1/x + \int_0^{\infty} e^{-t}(x-t)^{-2} dt \right)$ is the exponential
integral \cite{Ryzhik}, and $\gamma$ is Euler's
constant. Differentiation of Eq.(\ref{eq:specular}) in combination
with Eq.(\ref{eq:dF/dE-exp}) now yields:

\[
{{\partial I} \over {\partial E}} + \beta I + {{2\pi l^2} \over {A E}}
e^{-\beta E} \left[ \gamma + \log(\beta E) - {\rm Ei}(\beta E) \right]
= 0 ,
\]

\noindent which can be used to best-fit the correlation length and the
variance for a model of exponentially decaying correlations.

\section{Summary and Conclusions}
\label{conclusions}

This paper has 
presented new results on the inversion of structure of randomly
corrugated surfaces from atom scattering data, within the SA. This
work has been largely formal, with applications to be worked out in
the future in connection with presently unavailable experimental data.
The analysis presented here showed that
within the framework of the SA, the scattering intensities in
principle contain the full statistical information characterizing the
surface disorder. Several potentially useful expressions were derived
from which statistical parameters can be extracted in simple He
scattering experiments. One application of the theory presented here
is contained in our work on scattering from
Ag/Pt(111) \cite{me:Ag-systems}, in which it was shown how randomly
distributed adsorbates affect the scattering intensities. Additional
theoretical applications will be undertaken in the future, but it is
hoped above all that this work will stimulate experimentalists to
further utilize inert atom scattering in the study of increasingly
complex surface disorder. The results presented here suggest that such
experiments can reveal a wealth of information concerning disordered
surface structure, in particular on the statistics of randomly
corrugated surfaces.

\section*{Acknowledgements}

This work was carried out while the author was with the Physics
Department and the Fritz Haber Center for Molecular Dynamics at the
Hebrew University of Jerusalem, Givat Ram, Jerusalem 91904,
Israel. Numerous helpful discussions with Prof. R. Benny Gerber,
without whom this work could not have been completed, are gratefully
acknowledged. Partial support from NSF Grant CHE 97-32758 is gratefully
acknowledged as well.

\end{document}